\documentclass{svmult}

\usepackage{mathptmx}       
\usepackage{helvet}         
\usepackage{courier}        
\usepackage{makeidx}         
\usepackage{graphicx}        
\usepackage{multicol}        
\usepackage[bottom]{footmisc}

\usepackage[utf8]{inputenc}
\usepackage[T1]{fontenc}
\usepackage{amsmath}
\usepackage{amssymb}
\usepackage{mathrsfs}

\usepackage{microtype}
\usepackage{cite}

\newcommand*{\scri}{\mathscr{I}}
\newcommand*{\dd}{\mathrm{d}}

\newcommand*{\del}{\partial}

\newcommand*{\ba}{\mathbf{a}}
\newcommand*{\bb}{\mathbf{b}}
\newcommand*{\etp} {\eth'}
\newcommand*{\Y}[2]{{}_{#1}Y_{#2}}

\begin{document}

\title*{Linearized gravitational waves near space-like and null infinity}
\author{Florian Beyer \and George Doulis \and J\"org Frauendiener \and Ben Whale}
\institute{Florian Beyer \at Department of Mathematics and Statistics, University of
  Otago, P.O. Box 56, Dunedin 9010, New Zealand \email{fbeyer@maths.otago.ac.nz} \and
  George Doulis \at Department of Mathematics and Statistics, University of Otago,
  P.O. Box 56, Dunedin 9010, New Zealand \email{gdoulis@maths.otago.ac.nz} \and J\"org
  Frauendiener \at Department of Mathematics and Statistics, University of Otago, P.O. Box
  56, Dunedin 9010, New Zealand \email{joergf@maths.otago.ac.nz} \and Ben Whale \at
  Department of Mathematics and Statistics, University of Otago, P.O. Box 56, Dunedin
  9010, New Zealand \email{bwhale@maths.otago.ac.nz} }
%
%
\maketitle

\abstract{Linear perturbations on Minkowski space are used to probe numerically the remote
  region of an asymptotically flat space-time close to spatial infinity. The study is
  undertaken within the framework of Friedrich's conformal field equations and the
  corresponding conformal representation of spatial infinity as a cylinder. The system
  under consideration is the (linear) zero-rest-mass equation for a spin-2 field. The
  spherical symmetry of the underlying background is used to decompose the field into
  separate non-interacting multipoles. It is demonstrated that it is possible to reach
  null-infinity from initial data on an asymptotically Euclidean hyper-surface and that
  the physically important radiation field can be extracted accurately on $\scri^+$.}

 \section{Introduction}
 \label{sec:introduction}

 Asymptotically simple space-times as defined by Penrose~\cite{Penrose:1965wj} are
 distinguished by their qualitatively completely different structure at infinity. Depending
 on the sign of the cosmological constant $\lambda$ their conformal boundary is time-like,
 space-like or null. These space-times have been studied extensively from the point of view
 of the initial (boundary) value problem and a lot is known about their properties. In
 particular, space-times which are asymptotically de Sitter ($\lambda>0$) are known to
 exist globally for small initial data. The same is true for asymptotically flat
 space-times ($\lambda=0$) while for asymptotically anti-de Sitter space-times we only have
 short-time existence results for the initial boundary value problem. On the contrary,
 initiated by an important paper by Bizo\'n and Rostrowroski~\cite{Bizon:2011kz} and based
 on numerical and perturbative methods, there are now strong hints that these space-times
 are in fact non-linearly unstable (for more details on this topic see the contribution by
 O. Dias in this volume).

 In the present paper we want to focus on asymptotically flat space-times. As we already
 mentioned these space-times exist globally for small enough initial data.  However, it is
 still not completely understood in detail, how to characterize the admissible initial data
 from a geometrical or physical point of view. The reason lies in the fact that in every
 space-time with a non-vanishing ADM-mass spatial infinity is necessarily singular for the
 4-dimensional conformal structure.

 In a seminal paper~\cite{Friedrich:1998tc}, Friedrich shows how to set up a particular
 gauge near space-like infinity which exhibits explicitly the structure of the space-time
 near space-like and null-infinity. The so called `conformal Gau\ss\ gauge' is based
 entirely on the conformal structure of the space-time fixing simultaneously a coordinate
 system, an orthogonal frame and a general Weyl connection compatible with the conformal
 structure. The detailed description of this gauge is beyond the scope of this paper and we
 need to refer the reader to the existing literature~\cite{Friedrich:1995uf,
   Friedrich:1987ul, Friedrich:1998tc, Frauendiener:2004te}. Suffice it to say that the
 conformal Gau\ss\ gauge is based on a congruence of time-like conformal geodesics
 emanating orthogonally from an initial space-like hyper-surface. The coordinate system,
 tetrad and Weyl connection are defined initially on that hyper-surface and are dragged
 along the congruence of conformal geodesics so that they are defined everywhere.

 It turns out that in this gauge the space-time region near space-like infinity has a
 boundary consisting of future and past null-infinity $\scri^\pm$ together with a
 3-dimensional `cylindrical' hyper-surface $I$ connecting them. In a certain sense, in
 Minkowski space-time, this cylinder is a blow-up of the (regular) point $i^0$. The
 `general conformal field equations' (GCFE) express the fact that the conformal class of
 the space-time contains an Einstein metric. They are a set of geometric partial
 differential equations (PDEs) generalizing the standard Einstein equations with
 cosmological constant. When split into evolution and constraint equations within the
 conformal Gauß gauge the evolution equations take a particularly simple form and it turns
 out that the cylinder $I$ becomes a \emph{total characteristic}. This means that the
 evolution equations reduce to an intrinsic system of PDEs on $I$, i.e., they contain no
 derivatives transverse to~$I$. The intrinsic equations are symmetric hyperbolic on $I$
 except for the locations $I^\pm:=I\cap \scri^\pm$, the 2-spheres where $I$ meets future or
 past null-infinity. There, the equations loose hyperbolicity. It is this feature which is
 responsible for the singular behavior of the conformal structure near space-like infinity.

 In a series of papers~\cite{Friedrich:2008gs, Friedrich:1998tc, Friedrich:2008wt,
   Friedrich:2007vc}, Friedrich has analyzed the behavior of the fields, in particular of
 the Weyl tensor components and their transverse derivatives, along the cylinder $I$. He
 has shown that generic initial data lead to singularities at $I^\pm$. The singularities
 can be avoided if the initial data satisfy certain conditions, one of them being the
 geometric condition that the Cotton tensor of the induced geometry on the initial
 hyper-surface and all its symmetric trace-free derivatives vanish at the intersection of
 $I$ and the initial hyper-surface. It is still not completely clear what is the correct
 geometric classification of those space-times which satisfy Friedrich's conditions
 (however, see~\cite{Friedrich:2007vc,Friedrich:2008wt,Friedrich:2012vc}).

 Here, we want to discuss these issues from the numerical point of view. It is clear, that
 the fact that the evolution equations extracted from the GCFE cease to be hyperbolic is a
 troublesome feature in a numerical evolution scheme. We want to see how it manifests
 itself in the simplest scenario we can think of: linearized gravitational fields on
 Minkowski space. If we could not control this situation then the use of the GCFE for
numerical purposes would not be possible.

The plan of the paper is as follows: in Sect.~\ref{sec:spin-2-zero} and
Sect.~\ref{sec:numer-meth-tests} we give brief summaries of the basic analytical and
numerical results. In Sect.~\ref{sec:beyond-i+} we discuss ways to overcome the singular
behavior at $I^+$ while in Sect.~\ref{sec:reach-null-infin} we show how we can reach
future null-infinity. The conventions we use and much of the background  can be found in~\cite{Penrose:1984wm}.

 \section{The spin-2 zero-rest-mass equations}
 \label{sec:spin-2-zero}

 It is well known~\cite{Penrose:1984wm} that perturbations of the Weyl tensor $C_{abc}{}^d$
 or, equivalently, the Weyl spinor $\Psi_{ABCD}$ satisfy the equations for a field with
 spin~2 and vanishing rest-mass:
 \begin{equation}
   \label{eq:1}
   \nabla^A{}_{A'} \phi_{ABCD} = 0.
 \end{equation}
 It is also known~\cite{Penrose:1984wm} that this system of equations suffers from Buchdahl
 conditions, algebraic conditions relating the conformal curvature $\Psi_{ABCD}$ of the
 underlying background geometry with the field
 \[
 \Psi_{ABC(D} \phi^{ABC}{}_{E)} = 0
 \]
 which severely restrict the possible perturbations in any conformally curved space-time,
 rendering~\eqref{eq:1} inconsistent.

 Therefore, we choose flat Minkowski space-time as our background so that $\phi_{ABCD}$
 describes small amplitude gravitational waves propagating in an otherwise empty
 space-time. For a detailed discussion and derivation of the explicit form of the spin-2
 equations in the present context we refer to~\cite{Beyer:2012ie}. Here, we focus only on
 the relevant points. Starting with the standard Minkowski metric in Cartesian coordinates
 $X^\ba$
 \begin{equation}
   \label{eq:2}
   \tilde g = \eta_{\ba\bb} \dd X^\ba \dd X^\bb 
 \end{equation}
 where $\eta_{\ba\bb}=\mathrm{diag}(1,-1,-1,-1)$, and performing an inversion at the null-cone
 of the origin
 \[
 X^\ba = - \frac{x^\ba}{x\cdot x}, \qquad x\cdot x :=\eta_{\ba\bb}x^\ba x^\bb =
 \frac1{(X\cdot X)}
 \]
 puts the metric into the form
 \begin{equation}
   \label{eq:3}
   \tilde g = \frac{\eta_{\ba\bb}\dd x^\ba \dd x^\bb}{(x\cdot x)^2}.
 \end{equation}
 This metric is singular whenever $x\cdot x=0$, i.e., on the null-cone at infinity. So we
 define a conformally related metric
 \begin{equation}
   \label{eq:4}
   g' = \Omega^2 \tilde g = \eta_{\ba\bb}\dd x^\ba \dd x^\bb , \qquad \Omega = -(x\cdot x)
 \end{equation}
 which extends smoothly to the null-cone of infinity.  Note, that space-like infinity is
 represented in this metric as the point $x^\ba=0$. In order to exhibit the cylindrical
 structure referred to above we perform a further rescaling of the metric using a function
 $\kappa(r) = r \mu(r)$, where $r^2 = (x^1)^2 + (x^2)^2 + (x^3)^2$ and $\mu$ is a smooth
 function with $\mu(0)=1$. Furthermore, we introduce a new time coordinate by defining $x^0
 = t \kappa(r)$. These steps give the final form of the metric
 \begin{equation}
   \label{eq:5}
   g = \frac1{\kappa^2} \left(\kappa^2 \dd t^2 + 2 t \kappa \kappa' \dd t \dd r - (1-t^2 \kappa'{}^2) \dd r^2 - r^2 \dd \omega^2 \right).
 \end{equation}
 Here, we have denoted the metric on the unit sphere by $\dd\omega^2$. Note, that the
 metric $g$ is spherically symmetric.

\begin{figure}[htb]
   \centering
   \includegraphics[height=8cm]{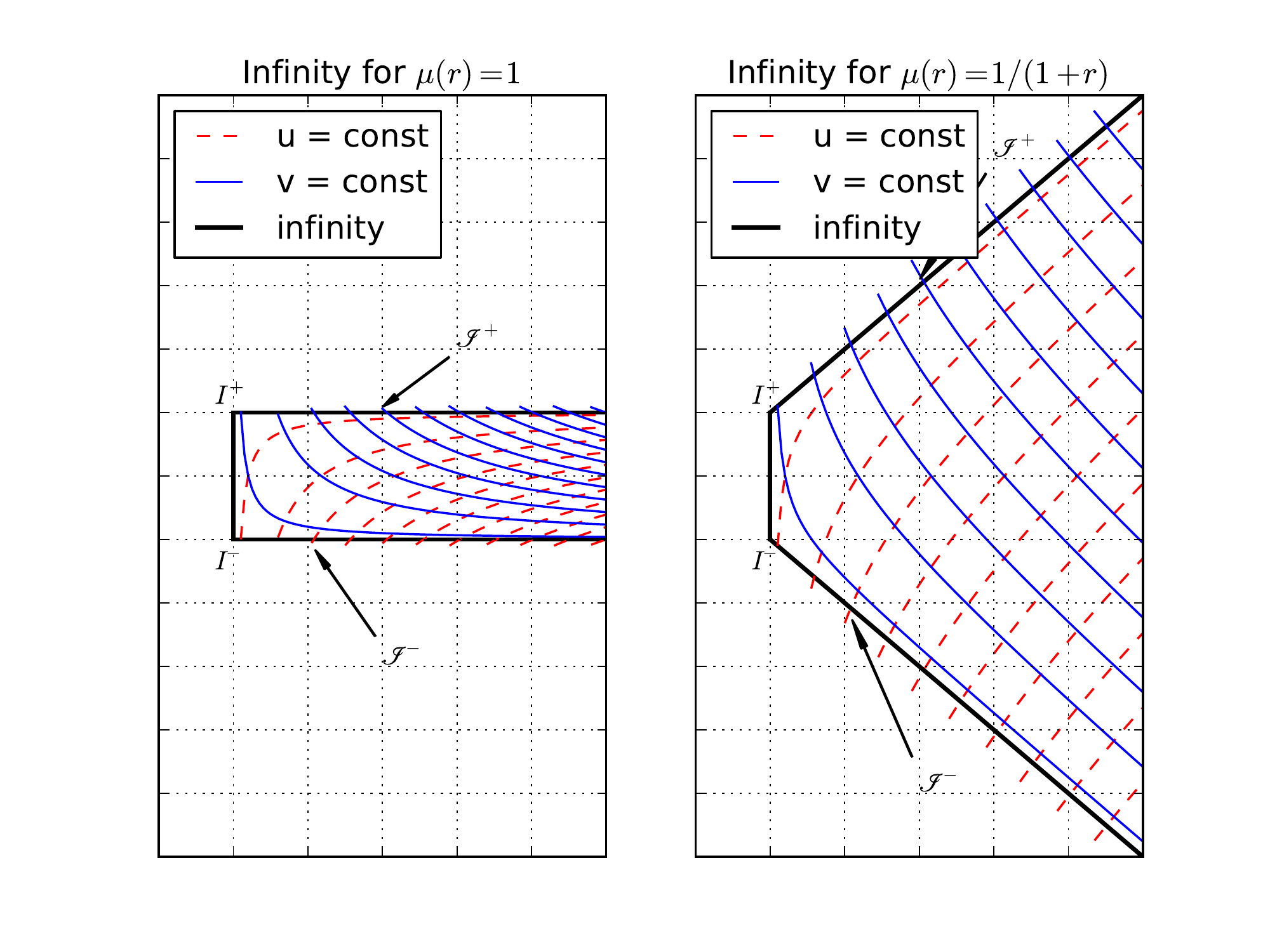}
   \caption{The neighborhood of space-like infinity in two different $(t,r)$-coordinate
     representations corresponding to $\mu(r)=1$ and $\mu(r)=1/(1+r)$}
   \label{fig:cylinder}
 \end{figure}

 The function $\mu(r)$ determines the `shape' of $\scri^\pm$ in the $(t,r)$-coordinate
 system. We have chosen the two possibilities $\mu(r)=1/(1+nr)$ with either $n=0$ or $n=1$
 with the consequence that $\scri^\pm$ are represented as either a horizontal ($n=0$) or
 diagonal ($n=1$) line in a $(t,r)$ diagram, see Fig.~\ref{fig:cylinder}. They meet the
 cylinder $I$ in the spheres $I^\pm$ at $(t,r)=(\pm1,0)$.
 We introduce the double null-coordinates $(u,v)$ by
 \begin{equation}
   \label{eq:6}
   u = \kappa t - r, \qquad v = \kappa t + r,
 \end{equation}
 which puts the metric into the form
 \[
 g = \frac1{\kappa^2}\dd u \dd v - \frac1{\mu^2}\dd\omega^2.
 \]
 Null-infinity is characterized by the vanishing of one of the null coordinates, the other
 one being non-zero ($u=0$, $v\ne0$ on $\scri^+$ and $u\ne0$, $v=0$ on $\scri^-$). Both
 coordinates vanish on the cylinder~$I$. In Fig.~\ref{fig:cylinder} we also display the
 lines of constant $u$ and $v$, corresponding to radial null geodesics. There are two
 families of null geodesics, one leaving the space-time through $\scri^+$, the other
 entering it through $\scri^-$.

 Since the null coordinates are adapted to the conformal structure, we can see clearly that
 the cylinder is `invisible' from the point of view of the conformal structure. This
 structure is exactly the same as the representation of space-like infinity in the
 coordinates $x^\ba$, it appears like a point. However, the differentiable structures defined
 by the $(u,v)$ and the $(t,r)$ coordinates are completely different near the
 boundary~$r=0$ from the one defined by the $x^\ba$ coordinates.

 In order to derive the spin-2 equations explicitly we introduce a complex null tetrad
 $(l^a, n^a, m^a, \bar m^a)$ adapted to the spherical symmetry, i.e.,
 \[
 l^a\del_a = \frac1{\sqrt2}\left(\left(1 - t \kappa'\right) \del_t + \kappa \del_r \right),
 \qquad n^a\del_a = \frac1{\sqrt2}\left(\left(1 + t \kappa'\right) \del_t - \kappa \del_r
 \right),
 \]
 and $m^a$ tangent to the spheres of symmetry.  Writing the spin-2 equation in the NP
 formalism, computing the spin-coefficients and finally introducing the `eth' operator
 $\eth$ (see~\cite{Goldberg:1967tf, Newman:1968tv, Penrose:1984wm}) on the unit sphere
 puts~\eqref{eq:1} into the form of eight coupled equations
 \begin{equation}
   \label{eq:7}
   \begin{aligned}
     (1-t\kappa') \del_t \phi_k + \kappa \del_r \phi_k - (3\kappa'  - (5-k) \mu) \phi_k &=  \mu\etp \phi_{k-1} , &k=1:4,\\
     (1+t\kappa') \del_t \phi_{k} - \kappa \del_r \phi_{k} + (3\kappa' + (k+1) \mu )
     \phi_{k} &= \mu \eth \phi_{k+1}, &k=0:3
   \end{aligned}
 \end{equation}
 for the five complex components of the spin-2 field $\phi_{ABCD}$, see
 e.g.~\cite{Penrose:1984wm}.  We exploit the spherical symmetry of the background
 Minkowski space-time even further by decomposing the field components
 $\phi_k(t,r,\theta,\phi)$ into different multipole moments using the spin-weighted
 spherical harmonics $\Y{s}{lm}$
 \begin{equation}
   \label{eq:8}
   \phi_k(t,r,\theta,\phi) = \sum_{l\ge 2-k} \sum_{m=-l}^l   \phi^{lm}_k(t,r)\; \Y{2-k}{lm}(\theta,\phi) .
 \end{equation}
 Inserting this expansion into~\eqref{eq:7} and using the action of $\eth$ and $\etp$ on
 the spin-weighted spherical harmonics \cite{Penrose:1984wm} the system decouples into a
 countable family of $1+1$ systems indexed by admissible pairs of integers $(l,m)$. We find
 that each mode $\phi_k^{lm}$ propagates along radial null geodesics, the `inner' modes
 $\phi_1^{lm}$, $\phi_2^{lm}$, $\phi_3^{lm}$ propagate in both directions while the `outer'
 modes $\phi_0^{lm}$ and $\phi_4^{lm}$ only propagate along one null direction.

 The final step in the preparation of the equations is the split into constraint and
 evolution equations, which leads to a system of five evolution equations
 \begin{equation}
   \label{eq:9}
   \begin{aligned}
     (1+t\kappa') \del_t \phi_0 - \kappa \del_r \phi_0 &= -(3\kappa' - \mu) \phi_0   -\mu \alpha_2 \phi_1 ,\\
     \del_t \phi_1  &=  -  \mu \phi_1 +  \frac12 \mu \alpha_2 \phi_0 - \frac12 \mu \alpha_0 \phi_2,\\
     \del_t \phi_2  &=    \frac12 \mu \alpha_0 \phi_1 - \frac12 \mu \alpha_0 \phi_3,\\
     \del_t \phi_3  &=   \mu \phi_3 +  \frac12 \mu \alpha_0 \phi_2 - \frac12 \mu \alpha_2 \phi_4,\\
     (1-t\kappa') \del_t \phi_4 + \kappa \del_r \phi_4 &= (3\kappa' - \mu) \phi_4 + \mu
     \alpha_2 \phi_3
   \end{aligned}
 \end{equation}
 and three constraint equations
 \begin{equation}
   \label{eq:10}
   \begin{aligned}
     - 2 \kappa \del_r \phi_1 + 6 r \mu' \phi_1 - 2 t \kappa' \mu \phi_1 +  \alpha_0 \mu (1-t\kappa') \phi_2 + \alpha_2 \mu (1+t\kappa') \phi_0 &= 0,\\
     -2 \kappa \del_r \phi_2 + 6 r \mu' \phi_2  + \alpha_0 \mu (1-t\kappa') \phi_3 + \alpha_0 \mu (1+t\kappa') \phi_1 &= 0,\\
     - 2 \kappa \del_r \phi_3 + 6 r \mu' \phi_3 + 2 t \kappa' \mu \phi_3 + \alpha_0 \mu
     (1+t\kappa') \phi_2 + \alpha_2 \mu (1-t\kappa') \phi_4 &= 0.
   \end{aligned}
 \end{equation}
 Note that we have dropped here the superscripts from $\phi_k^{lm}$ and we introduced the
 quantities $\alpha_0 = \sqrt{l(l+1)}$ and $\alpha_2=\sqrt{l(l+1)-2}$. Since $\kappa(0)=0$
 it is obvious that the evolution equations reduce to a system intrinsic to $I$. Since on
 $I$ also $\kappa'(0)=1$, it follows that the coefficients in front of the time
 derivatives of $\phi_0$ (respectively $\phi_4$) vanish when $t=-1$ (respectively when
 $t=1$).

 \begin{figure}[htb]
   \centering
   \includegraphics[width=0.7\textwidth]{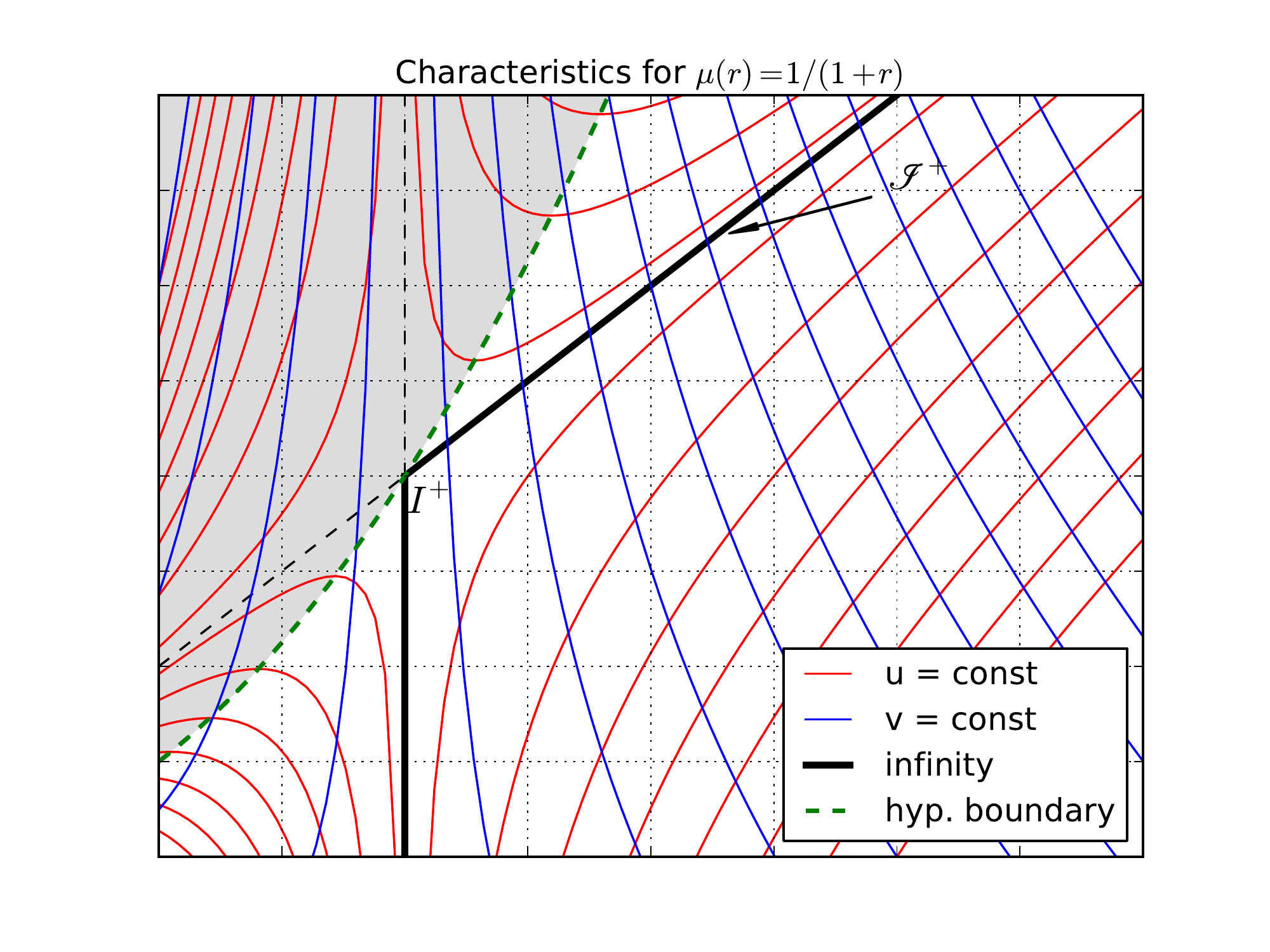}
   \caption{The characteristics of the evolution equations in a neighborhood of $I^+$ in
     the diagonal representation. The shaded region is the domain where the equations fail
     to be hyperbolic, i.e., where $t<\kappa'(r)$. The corresponding neighborhood of $I^-$
     is obtained by reflection at the lower border of the diagram accompanied by an
     interchange of $u$ and $v$. The thick (green) broken line indicates the boundary of
     the domain of non-hyperbolicity.}
   \label{fig:2}
 \end{figure}
 In Fig.~\ref{fig:2} we show again the neighbourhood of $I$ and $\scri^+$ for
 $\mu(r)=1/(1+r)$. This time we show the characteristics also in the unphysical part of
 the diagram. The non-shaded region bounded partly by the thick broken line is the domain
 of hyperbolicity of the evolution equations, i.e., the domain where $t<\pm
 |\kappa'(r)|$. Notice that every neighborhood of $I^+$ contains regions where the
 hyperbolicity breaks down. Apart from the fact that one cannot hope to get existence and
 uniqueness of solutions beyond that region its presence also makes the numerical
 evolution challenging. For instance, setting up an initial boundary value problem with
 the left boundary at negative values of $r$ does not make sense if the evolution is to
 reach up to $I^+$. Even with the left boundary on $I$ it is not possible to go beyond
 $I^+$ since the evolution hits the non-hyperbolicity region.

 \section{Numerical methods and tests}
 \label{sec:numer-meth-tests}

 Again, we give a brief summary and refer to~\cite{Beyer:2012ie} for further details. We
 have different equations for different values of the multipole index $l$. Here, we focus
 only on the case $l=2$ without mentioning it any further. We have done evolutions with
 other values of $l$ which lead to very similar results. The
 evolution equations~\eqref{eq:9} are solved on a spatial interval $0 \le r \le 1$ as an
 initial boundary value problem. We use the method of lines constructing the spatial
 discretisation using a $4^\text{th}$ order accurate finite difference scheme.  The initial
 data are obtained either from exact solutions or by solving the constraint
 equations~\eqref{eq:10} explicitly in terms of two free functions. The boundary at $r=0$
 is a characteristic so we do not need to specify any free functions there, while the
 boundary at $r=1$ is an artificial boundary, which needs exactly one free function for the
 component $\phi_0$ that propagates through this boundary into the computational
 domain. The boundary conditions are implemented using SBP operators \cite{Strand:1994ef,
   Gustafsson:1995vp, Diener:2007bx} and the SAT penalty method~\cite{Carpenter:1994cu,
   Carpenter:1999cl, Lehner:2005hc, Schnetter:2006ku}. The semi-discrete system of ODEs is
 solved using the standard $4^\text{th}$ order Runge-Kutta method. 

 This code stably propagates the initial data from $t=0$ up to any value of $t < 1$,
 independently of whether we use the diagonal or the horizontal representation of
 $\scri$. The code converges to $4^\text{th}$ order, the constraints propagate and the
 constraint violations remain bounded. However, the two representations behave differently
 when we attempt to reach $t=1$.

 In the diagonal case we can reach $t=1$ exactly and the code converges in $4^\text{th}$
 order, indicating that the numerical problem is still well-posed. With any fixed time-step
 $\tau$ we can also step beyond $t=1$ to $t=1+\tau$. However, the code fails to converge at
 $t=1+\tau$ for every $\tau>0$. Clearly, the loss of hyperbolicity at $t=1$ is also
 responsible for the lack of well-posedness of the numerical problem. Referring back to
 Fig.~\ref{fig:2} it is clear that there will always be a sufficiently high resolution in
 time and space which will detect the domain of non-hyperbolicity of the evolution
 equations beyond $I^+$.

 In the horizontal case the situation is apparently worse. In this case, it is
 \emph{principally impossible} to reach $t=1$ simply because the propagation speed of
 $\phi_4$ becomes infinite and so stability is lost not only at the point $I^+$ but at the
 entire time-slice $t=1$ which coincides with $\scri^+$, a characteristic hyper-surface, in
 this representation.

 However, \emph{in principle} we can come arbitrarily close to $t=1$ (within numerical
 accuracy) if at every time we choose the next time-step small enough so that the CFL
 criterion---the numerical domain of dependence includes the analytical domain of
 dependence---is satisfied. Such an adaptive time-stepping scheme has been implemented and
 we have in fact demonstrated that we can come very close to $t=1$ without loosing
 convergence. However, since the time-steps decrease to zero exponentially with the number
 of steps the simulation takes arbitrarily long.

 \section{Beyond $I^+$?}
 \label{sec:beyond-i+}

 Given that we can reach $t=1$ in a stable fashion in the diagonal representation we may
 ask the question as to whether it is possible to continue beyond $I^+$? Clearly, for the
 reasons discussed above we cannot simply continue the computation because we run into the
 non-hyperbolicity region near $I^+$.  Referring back to Fig.~\ref{fig:2} we see that one
 family of the characteristics---corresponding to the field component $\phi_4$---asymptotes
 to the cylinder and, subsequently, to $\scri^+$ in a non-uniform way. This non-uniformity
 is due to the fact that the coefficient in front of the time derivative of $\phi_4$ in the
 evolution system vanishes at $t=1$, causing the loss of hyperbolicity of the system.  Can
 we avoid this problem? There are a few possibilities which come to mind:
 \begin{itemize}
 \item Chop the computational domain by dropping points outside $\scri^+$ from the left.
 \item Change the radial coordinate so that $\scri^+$ becomes the left boundary of the
   computational domain. This is a form of $\scri$-freezing
   (see~\cite{Frauendiener:1998th}).
 \item Change the time coordinate so that the (space-like) time-slices tilt upwards
   towards~$\scri^+$.
 \end{itemize}
 We have considered the first two possibilities. These two methods are complementary to
 each other in the sense that in the first case we change the computational domain but not
 the system, while in the second case we change the equations but not the computational
 domain.

 Fig.~\ref{fig:2} shows the behavior of the characteristics for $\phi_0$ and $\phi_4$ as
 well as the region in which hyperbolicity fails. The surfaces that we are evolving our
 data on are horizontal and to the right of the vertical solid and broken black lines. Note
 that for $t>1$ these surfaces are guaranteed to intersect the region in which
 hyperbolicity fails.  In addition, for surfaces with $t>1$ a new boundary condition for
 $\phi_4$, on the `left' of the grid is required. However, since $\scri^+$ is one of the
 characteristics for $\phi_4$ this boundary condition cannot influence the physical region
 to the right of $\scri^+$.

 To solve both of these problems we set the values of $\phi_0,\ldots,\phi_4$ to zero after
 some predetermined grid point that is beyond future null infinity, but before the region
 in which hyperbolicity fails. That is, between the broken green line and $\scri^+$.  The
 equation for the broken green line is
 \[
 1 - \frac{t}{(r+1)^2} = 0
 \]
 and the equation for future null infinity, where $t\geq 1$, is
 \[
 1 - \frac{t}{r+1}= 0.
 \]
 We chose, after some experimentation, that the `cut off' point, after which all data
 values would be set to zero would be $10\%$ of all grid points beyond $\scri^+$. That is,
 the values of $\phi_0,\ldots,\phi_4$ would be set to zero on the $90\%$ of grid points
 between $r=0$ and $\scri^+$. This condition, of setting values to zero, allows us to
 numerically provide the necessary boundary condition for $\phi_4$ and cope with the lack
 of hyperbolicity in the region of the grid to the left of the broken green line.

 Other than this technicality, the same evolution scheme, with the same fixed step size was
 used to evaluate the initial data up to $t=1.1$. We evolved an exact solution and
 estimated the error at $t=1.1$ by comparing the values of $\phi_0,\ldots,\phi_4$, on the
 `physical' portion of the grid to the exact solution. The $L_2$ measure of the error is
 presented in Table~\ref{tab:n1L2err}. We only give the error for the $\phi_0$ and $\phi_4$
 components. The convergence rates for $\phi_1,\phi_2$ and $\phi_3$ are, roughly, a linear
 interpolation between those for $\phi_0$ and $\phi_4$. Note, that the error in $\phi_0$ is
 of the order of $10^{-11}$ while the error in $\phi_4$ is roughly $10^{-2}$.

 \begin{table}[htb]
   \centering
   \begin{tabular}{c||c|r|c|r|}
     & \multicolumn{2}{c|}{$\phi_0$} & \multicolumn{2}{c|}{$\phi_4$}\\
     Grid Points& $\log_2(||\Delta||_2)$ & Rate 
     & $\log_2(||\Delta||_2)$ & Rate \\\hline\hline
     100 & -28.38 &      & -5.93  &  \\
     200 & -31.75 & 3.37 & -5.98  & 0.05 \\
     400 & -35.21 & 3.46 & -6.07  & 0.10 \\
     800 & -38.71 & 3.50 & -8.41  & 2.34 
   \end{tabular}
   \caption{Absolute error $\Delta$ compared to an exact solution.
     $L^2$ norm and convergence rates at time $t=1.1$ for $\phi_0, \phi_4$. 
     The calculation was done with a fixed time-step.}\label{tab:n1L2err}
 \end{table}

 In the second case we change the $r$-coordinate in a very simple minded way using the
 coordinate transformation
 \[
 r\mapsto \bar r = r-t+1, \qquad t \mapsto \bar t = t, \qquad \text{for } t\ge 1.
 \]
 This has the effect that $\scri^+$ is given as the locus $\{\bar r =0\}$. Note, that the
 coordinate transformation is only continuous and not even $\mathcal{C}^1$. Since the
 partial derivatives transform according to
 \[
 \del_t = \del_{\bar t} - \del_{\bar{r}}, \qquad \del_r = \del_{\bar{r}}
 \]
 we still have the problem that the coefficient in front of the time derivative of $\phi_4$
 vanishes at $t=1$. Clearly, we cannot make the system regular at $t=1$ by a coordinate
 transformation. We can try to avoid the evaluation at $t=1$ by arranging the time stepping
 to `straddle' $t=1$ so that we never actually hit it exactly. However, this has the
 consequence that the `kink' that we obtain due to the non-smoothness of the coordinate
 transformation induces oscillations at the right boundary. We would probably be able to
 avoid those if we used a smoother coordinate transformation. However, this would change
 the equations in much more complicated ways and we have not pursued this any further.

 The third possibility mentioned above---changing the time coordinate---has the effect that
 the time-slices globally approach $\scri^+$ which is exactly the feature that we see in
 the horizontal representation. This prompted us to look at the relationships between the
 different conformal representations in more detail.

 The two representations are related by a conformal rescaling and a coordinate
 transformation. We can derive the corresponding relationships for the field components as
 follows. Let $\mu_n(r) = 1/(1+n r)$ and define $\Theta = \mu_1/\mu_0$. Let $g_0$ and $g_1$
 be the metrics corresponding to the horizontal and diagonal representations, then we have
 \begin{equation}
   \label{eq:11}
   g_1 = \frac{\mu_1^2}{\mu_0^2}\, g_0 = \Theta^2 g_0.
 \end{equation}
 Furthermore, the two time coordinates $t_0$ and $t_1$ in the two representations are
 related by
 \begin{equation}
   \label{eq:12}
   x^0 = r \mu_0 t_0 = r \mu_1 t_1 \implies t_1 = \Theta^{-1} t_0.
 \end{equation}
 The spin-2 field has conformal weight $-1$ under conformal rescalings. This implies that
 with~\eqref{eq:11} we also have
 \begin{equation}
   \label{eq:13}
   \phi^1_{ABCD} = \Theta^{-1} \phi^0_{ABCD}.
 \end{equation}
 In order to get the behavior of the field components under conformal rescalings we observe
 that~\eqref{eq:11} implies that the tetrad vectors rescale as
 \begin{equation}
   \label{eq:14}
   l^a_1 = \Theta^{-1} l^a_0, \quad \text{etc}
 \end{equation}
 and the spin-frame correspondingly rescales with $\Theta^{-1/2}$. Taken altogether, these
 transformation properties imply that
 \begin{equation}
   \phi^1_k(t_1,r) = \Theta^{-3} \phi^0_k(\Theta^{-1} t_0,r).\label{eq:15}
 \end{equation}
 We can now make use of these relationships in the following way. Suppose we want to evolve
 initial data in the diagonal representation. We rescale the data on the initial
 hyper-surface $t_1=0=t_0$ using~\eqref{eq:15} into initial data for the horizontal
 representation and evolve them with the system for the horizontal representation up to a
 time $t\approx1$. Then we undo the rescaling with~\eqref{eq:15} and obtain the solution in
 the diagonal representation.
 \begin{figure}[htb]
   \centering
   \includegraphics[width=0.9\textwidth]{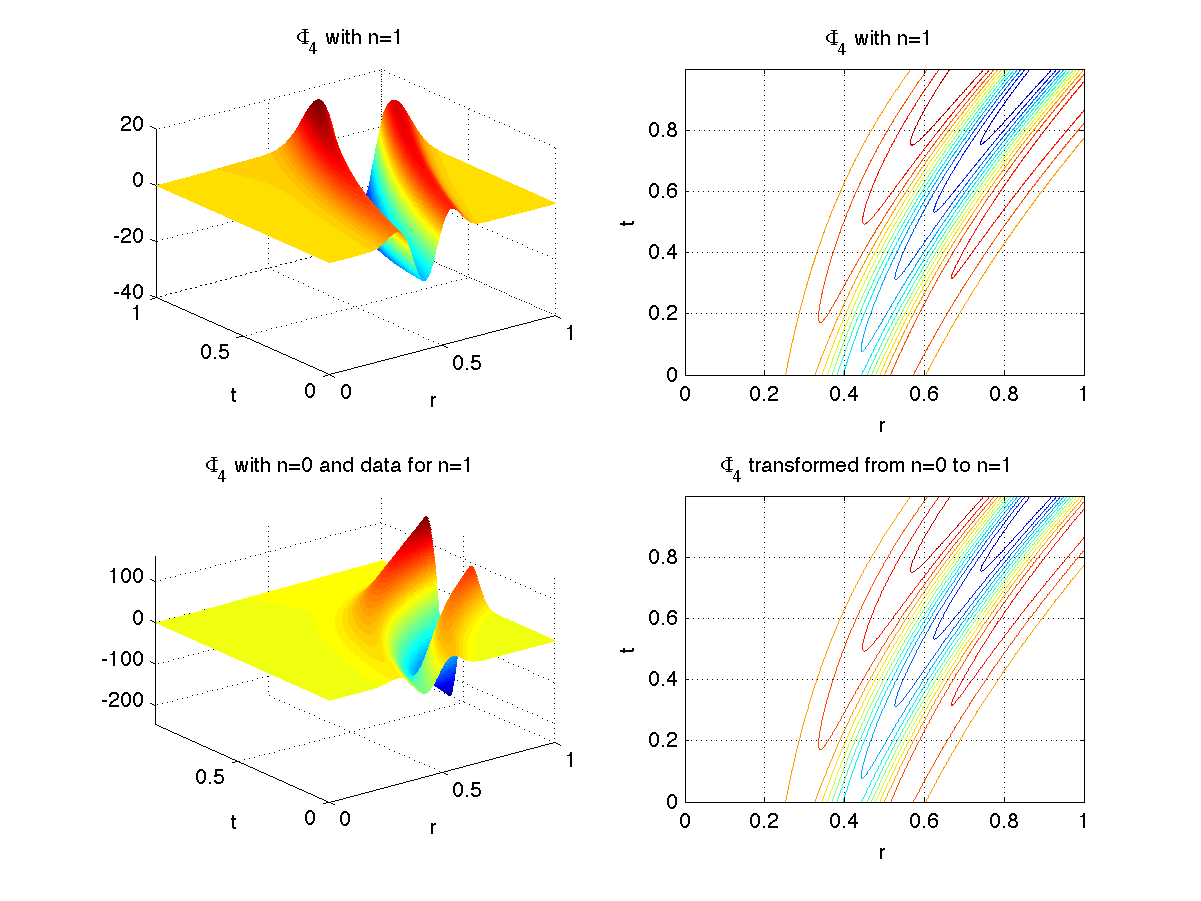}
   \caption{Comparison between the two representations. For detailed explanation see text.}
   \label{fig:phi4comparison}
 \end{figure}
 In Fig.~\ref{fig:phi4comparison} we present the results of these operations. The surface
 plot in the upper left shows the component $\phi_4$ evolved with the diagonal
 representation $n=1$, while the plot on the lower left shows $\phi_4$ obtained in the
 horizontal representation using the same data as for the case $n=1$ but rescaled
 according to~\eqref{eq:15}. The contour plots on the right show the obtained
 solutions. Above is the solution directly obtained in the diagonal representation while
 below is the solution obtained after rescaling back from the horizontal
 representation. The contour plots agree visually. Note, that the $t$ coordinates in the
 two surface plots are not the same. They refer to $t_0$ in the lower plot and to $t_1$ in
 the other.

\begin{figure}[htb]
   \centering
   \includegraphics[width=0.48\textwidth]{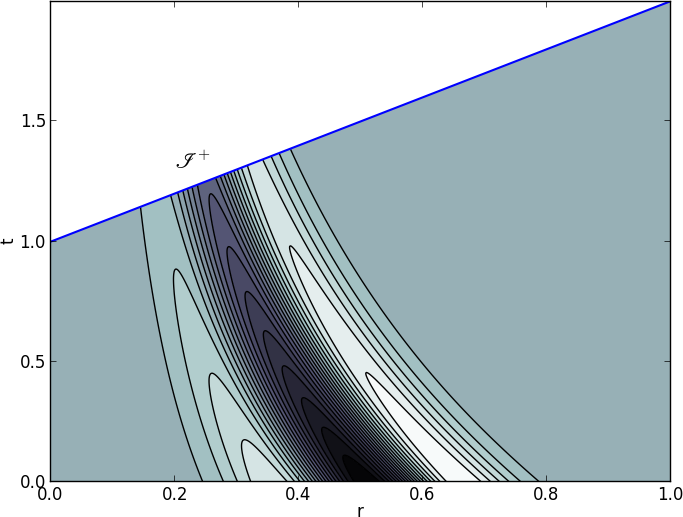}
   \hfill
   \includegraphics[width=0.48\textwidth]{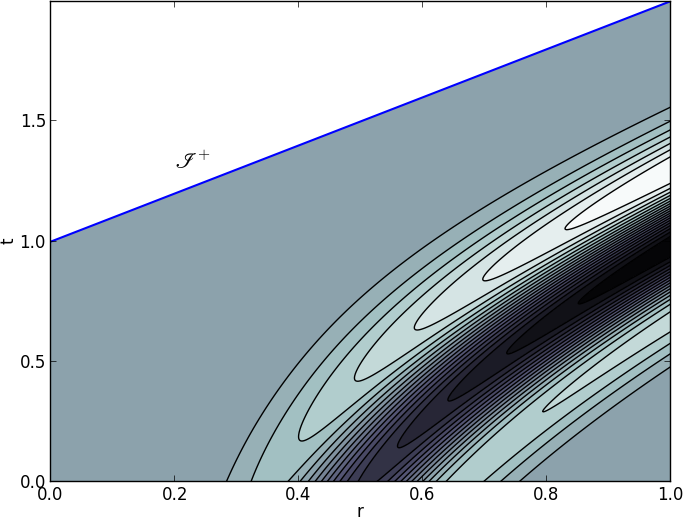}
   \caption{Contour plots of $\phi_0$ (left) and $\phi_4$ (right) computed with $n=0$ but
     rescaled to the diagonal representation.}
   \label{fig:phifull}
 \end{figure}

 In the horizontal representation we come arbitrarily close to $t=1$. Hence, we can
 `almost' compute the entire space-time (at least in this simple set-up). In this sense the
 horizontal representation is much more efficient than the diagonal one. In
 Fig.~\ref{fig:phifull}
 we show the contour plots of the rescaled solution in the diagonal
 representation. Clearly, it extends way over the $t=1$ time-slice reaching $\scri^+$ to
 within the graphical resolution.

 \section{Reaching null-infinity}
 \label{sec:reach-null-infin}

 Since the horizontal representation allows us to almost reach null-infinity one may wonder
 whether it might be possible to extend the computation to $\scri^+$ from the last
 time-level at $t=\bar{t}$, say? To answer this question we refer back to
 Fig.~\ref{fig:cylinder} where we showed the space-time in the horizontal representation
 together with the two families of characteristics. The `outgoing' characteristics (those
 intersecting $\scri^+$) are well-behaved while the other family asymptotes to
 $I\cup\scri^+$ in a non-uniform way. Referring to Eq.~\eqref{eq:7} we see that all
 components except for $\phi_4$ propagate along the well-behaved characteristics. This
 suggests that we use those four propagation equations to estimate the values of $\phi_0$,
 $\phi_1$, $\phi_2$, $\phi_3$ on $\scri^+$. Since we cannot use any values of $\phi_4$
 beyond $t=\bar{t}$ we are forced to use an Euler step to get to $t=1$.

 As for $\phi_4$ we observe that its propagation equation reduces to an intrinsic equation
 on $\scri^+$, which we could integrate if we had initial conditions for $\phi_4$ at $r=0$,
 once we know the values of the other components on $\scri^+$. The equation for $\phi_4$ on
 $\scri^+$ reduces to
 \begin{equation}
   \label{eq:16}
   r\del_r\phi_4 - 2 \phi_4 = \alpha_2 \phi_3.
 \end{equation}
 This equation is singular at $r=0$ and the requirement that $\phi_4$ and its derivative be
 bounded forces the initial condition
 \[
 \phi_4(1,0) = -\frac{\alpha_2}{2} \phi_3(1,0).
 \]
 So we see that we can perform the last step to $\scri^+$, albeit only with a first order
 method.

 We tested this approach by evolving compactly supported initial data obtained by explicitly
 solving the constraints (the same ones as in Sec.~\ref{sec:beyond-i+}). The error at $t=1$ was estimated by comparing the values produced in some
 simulation to the values produced in the highest resolution simulation with $6400$ grid
 points.  The $L_2$ measure of the error is presented in Table \ref{tab:n0L2err}.  Since
 the $\phi_0$ and $\phi_4$ components mark the extreme cases, we do not present the error
 convergence rates for $\phi_1,\phi_2,\phi_3$. Their convergence rates are comparable to $\phi_0$.
 \begin{table}[htb]
   \centering
   \begin{tabular}{c||c|r|c|r|}
     & \multicolumn{2}{c|}{$\phi_0$} & \multicolumn{2}{c|}{$\phi_4$}\\
     Grid Points& $\log_2(||\Delta||_2)$ & Rate & $\log_2(||\Delta||_2)$ & Rate \\\hline\hline
     200 & -7.99 &   & 0.082 &  \\
     400 & -11.48 & 3.49 & -0.44  & 0.52 \\
     800 & -14.98 & 3.49 & -1.02  & 0.57 \\
     1600 & -18.48 & 3.50 & -1.73 & 0.71 \\
     3200 & -22.07 & 3.58 & -2.81 & 1.08
   \end{tabular}
   \caption{Absolute error $\Delta$ compared to a 6400 grid point
     simulation using the 
     $L^2$ norm and convergence rates at time $t=1$ for $\phi_0, \phi_4$
     for compactly supported data in the $n=0$ space-time.}\label{tab:n0L2err}
 \end{table}
 Remarkably, the convergence rate for $\phi_0$ is close to $4^{\text{th}}$ order, while
 for $\phi_0$ we get the expected first order convergence. The accuracy in the radiation
 component is roughly $10^{-7}$ while for $\phi_4$ it is not good, only $\approx
 10\%$. This low accuracy is due to the singular behavior of the propagation
 equation~\eqref{eq:16} for $\phi_4$ at $r=0$, which we might not have taken care of
 appropriately. There is still room for improvement because we could use the intrinsic
 transport equations (see \cite{Beyer:2012ie,Friedrich:2003hp}) along the
 cylinder---mentioned in the introduction---to drag along as many derivatives of $\phi_4$
 as we please. These could be used to construct a Taylor approximation for $\phi_4$ near
 $r=0$ to get the integration process started more smoothly.

 \section{Conclusion}
 \label{sec:conclusion}

 In this paper we have presented a study of the spin-2 equation in the neighborhood of
 spatial infinity in Minkowski space-time. Since the perturbations of the Weyl curvature on
 flat space obey this equation we can interpret this system as a model for small amplitude
 gravitational waves. We used this model to study the asymptotic properties of the fields
 ---and, hence, to some extent also of the perturbed space-time---close to spatial infinity. This
 region is still not completely understood and we hope that our work will ultimately
 contribute to the complete understanding of this issue.

 We have shown here that it is possible to generate a complete evolution from initial data
 on an asymptotically Euclidean hyper-surface to the asymptotic regime including
 null-infinity even though the equations show a certain degeneracy at the cylinder which
 represents spatial infinity in Friedrich's conformal Gauß gauge. Using the horizontal
 representation as described in Sects.~\ref{sec:beyond-i+} and \ref{sec:reach-null-infin}
 we can get the radiation field $\phi_0$ quite accurately on $\scri^+$. The drop in
 convergence for the component $\phi_4$ is akin to the loss of smoothness of null-infinity
 in the first result on the global stability of Minkowski space by Christodoulou and
 Klainerman \cite{Christodoulou:1993vm}. This was caused by the loss of peeling in the
 corresponding component of the Weyl tensor.

 Of course, we deal here with a highly simplified system and the fact that we can make 
 things work here does not immediately imply that it will also work in the physically
 relevant 3D cases. However, this toy system does provide valuable insights into the
 possibilities of and the restrictions imposed by the structure of spatial infinity. We
 should point out that the structure of the fully non-linear general conformal field
 equations is very similar to the linear spin-2 system. This is essentially due to the fact
 that the spin-2 system results from the Bianchi equations obeyed by the rescaled Weyl spinor.

 Our next steps will be the removal of the artificial boundary at $r=1$. This will allow us
 to perform an entirely global evolution of the mode decomposed spin-2 field. The challenge
 here is to get the centre under control because due to the spherical symmetry this will be
 a singular point for the equations. Then, we intend to remove the mode decomposition and
 look at the linearized system in three spatial dimensions.

 \begin{acknowledgement}
   This research was supported in part by Marsden grant UOO0922 from the Royal Society of
   New Zealand. JF wishes to thank the organizers of the ERE2012 meeting in Guimar\~aes for
   their support.
 \end{acknowledgement}
 %


\begin{thebibliography}{10}
\providecommand{\url}[1]{#1}
\providecommand{\urlprefix}{URL }
\expandafter\ifx\csname urlstyle\endcsname\relax
  \providecommand{\doi}[1]{DOI~\discretionary{}{}{}#1}\else
  \providecommand{\doi}{DOI~\discretionary{}{}{}\begingroup
  \urlstyle{rm}\Url}\fi

\bibitem{Beyer:2012ie}
Beyer, F., Doulis, G., Frauendiener, J.: {Numerical space-times near space-like
  and null infinity. The spin-2 system on Minkowski space}.
\newblock Classical and Quantum Gravity \textbf{29}(24), 245,013 (2012)

\bibitem{Bizon:2011kz}
Bizo{\'n}, P., Rostworowski, A.: {On weakly turbulent instability of anti-de
  Sitter space}.
\newblock Phys. Rev. Lett. \textbf{107} (2011)

\bibitem{Carpenter:1994cu}
Carpenter, M.H., Gottlieb, D., Abarbanel, S.: {Time-stable boundary conditions
  for finite-difference schemes solving hyperbolic systems: methodology and
  application to high-order compact schemes}.
\newblock J. Comp. Phys. \textbf{111}(2), 220--236 (1994)

\bibitem{Carpenter:1999cl}
Carpenter, M.H., Nordstr{\"o}m, J., Gottlieb, D.: {A stable and conservative
  interface treatment of arbitrary spatial accuracy}.
\newblock J. Comp. Phys. \textbf{148}(2), 341--365 (1999)

\bibitem{Christodoulou:1993vm}
Christodoulou, D., Klainerman, S.: {The global nonlinear stability of the
  Minkowski space}, vol.~41.
\newblock Princeton University Press, Princeton (1993)

\bibitem{Diener:2007bx}
Diener, P., Dorband, E., Schnetter, E., Tiglio, M.: {Optimized high-order
  derivative and dissipation operators satisfying summation by parts, and
  applications in three-dimensional multi-block evolutions}.
\newblock J Sci Comput \textbf{32}(1), 109--145 (2007)

\bibitem{Frauendiener:1998th}
Frauendiener, J.: {Numerical treatment of the hyperboloidal initial value
  problem for the vacuum Einstein equations. II. The evolution equations}.
\newblock Phys Rev D \textbf{58}(6), 064,003 (1998)

\bibitem{Frauendiener:2004te}
Frauendiener, J.: {Conformal infinity}.
\newblock Living Rev. Relativity \textbf{7}, 2004--1, 82 pp. (electronic)
  (2004)

\bibitem{Friedrich:1995uf}
Friedrich, H.: {Einstein equations and conformal structure: Existence of
  anti-de Sitter-type space-times}.
\newblock J. Geom. Phys. \textbf{17}, 125--184 (1995)

\bibitem{Friedrich:1998tc}
Friedrich, H.: {Gravitational fields near space-like and null infinity}.
\newblock J. Geom. Phys. \textbf{24}, 83--163 (1998)

\bibitem{Friedrich:2003hp}
Friedrich, H.: {Spin-2 fields on Minkowski space near spacelike and null
  infinity}.
\newblock Classical and Quantum Gravity \textbf{20}(1), 101 (2003)

\bibitem{Friedrich:2007vc}
Friedrich, H.: {Static Vacuum Solutions from Convergent Null Data Expansions at
  Space-Like Infinity}.
\newblock Annales de l'IHP A  (2007)

\bibitem{Friedrich:2008gs}
Friedrich, H.: {Conformal classes of asymptotically flat, static vacuum data}.
\newblock Classical and Quantum Gravity \textbf{25}(6), 065,012 (2008)

\bibitem{Friedrich:2008wt}
Friedrich, H.: {One-parameter families of conformally related asymptotically
  flat, static vacuum data}.
\newblock Classical and Quantum Gravity  (2008)

\bibitem{Friedrich:2012vc}
Friedrich, H.: {Conformal structures of static vacuum data}.
\newblock arXiv \textbf{gr-qc} (2012)

\bibitem{Friedrich:1987ul}
Friedrich, H., Schmidt, B.G.: {Conformal Geodesics in General Relativity}.
\newblock Proc. Roy. Soc. A  (1987)

\bibitem{Goldberg:1967tf}
Goldberg, J.N., Macfarlane, A., Newman, E.T.: {Spin-s Spherical Harmonics and
  Eth}.
\newblock J Math Phys  (1967)

\bibitem{Gustafsson:1995vp}
Gustafsson, B., Kreiss, H.O., Oliger, S.: {Time-dependent problems and
  difference methods}.
\newblock A Wiley-Interscience Publication (1995)

\bibitem{Lehner:2005hc}
Lehner, L., Reula, O., Tiglio, M.: {Multi-block simulations in general
  relativity: high-order discretizations, numerical stability and
  applications}.
\newblock Classical and Quantum Gravity \textbf{22}(24), 5283 (2005)

\bibitem{Newman:1968tv}
Newman, E.T., Penrose, R.: {New conservation laws for zero rest-mass fields in
  asymptotically flat space-time}.
\newblock Proc. Roy. Soc. A \textbf{305}(1482), 175--204 (1968)

\bibitem{Penrose:1965wj}
Penrose, R.: {Zero rest-mass fields including gravitation: asymptotic
  behaviour}.
\newblock Proc. Roy. Soc. London A \textbf{284}, 159--203 (1965)

\bibitem{Penrose:1984wm}
Penrose, R., Rindler, W.: {Spinors and Spacetime: Two-spinor calculus and
  relativistic fields}, vol.~1.
\newblock Cambridge University Press, Cambridge (1984)

\bibitem{Schnetter:2006ku}
Schnetter, E., Diener, P., Dorband, E.N., Tiglio, M.: {A multi-block
  infrastructure for three-dimensional time-dependent numerical relativity}.
\newblock Classical and Quantum Gravity \textbf{23}(16), S553--S578 (2006)

\bibitem{Strand:1994ef}
Strand, B.: {Summation by Parts for Finite Difference Approximations for d/dx}.
\newblock J. Comp. Phys. \textbf{110}(1), 47--67 (1994)

\end{thebibliography}

\end{document}